\documentclass[aps,prd, 10pt, twocolumn,nofootinbib,showpacs]{revtex4}
\usepackage{subeqn}
\usepackage[usenames,dvipsnames]{xcolor}

\begin{document}

\title{Second order variations and the overspinning problem in Kerr-Taub-NUT space-time}
\author{Koray D\"{u}zta\c{s}}
\email{koray.duztas@okan.edu.tr}
\affiliation{\.{I}stanbul Okan University, Faculty of Engineering and Natural Sciences, Tuzla, \.{I}stanbul, 34959, T\"{u}rkiye }

\begin{abstract}
Previously, we attempted to over-spin Kerr-Taub-NUT black holes with test fields satisfying the null energy condition. We derived that a non-generic over-spinning could be achieved with strict restrictions on the test field modes and the initial space-time parameters. Here, we incorporate the effect of the second order variations by evaluating the Sorce-Wald condition for Kerr-Taub-NUT black holes. We calculate the second variation of the area for the sub-collection of Kerr-Taub-NUT black holes for which over-spinning is possible due to first order perturbations. We carry out an analysis to second order, avoiding the common order of magnitude problems where the second order variations appear in the form of a product with the square of an additional small parameter. We derive that the contribution of the second order variations fix the final state of the space-time parameters driving the Kerr-Taub-NUT black hole away from extremality. The event horizon is preserved in accord with the cosmic censorship conjecture and the laws of black hole dynamics.
\end{abstract} 
\maketitle
\section{Introduction}
\label{int}
Following the derivation of the Taub and its extension Taub-NUT metrics \cite{taub,nut}, Demia\'{n}ski and Newman presented the rotating extension known as the  Kerr-Taub-NUT space-time ~\cite{dem}. The Kerr-Taub-NUT (KTN) metric is an analytic solution of the Einstein field equations in vacuum endowed with three parameters namely: the gravitational mass $m$, the gravitomagnetic monopole moment or the NUT charge $- \ell$ and the Kerr parameter $a$. In addition to the usual singularities characterized by the divergence of the metric components, the KTN metric admits a second type of singularity which occurs when the determinant of the metric vanishes. This property is inherited from its predecessor Taub-NUT space-time. For the Taub-NUT case, Misner attempted to remedy the singularities of the second type at $\theta=0,\pi$ by imposing a periodicity on the time coordinate \cite{misner}. The same identification was adopted by Miller in a global analysis of the KTN space-time \cite{miller}. Imposing a periodicity condition on the time coordinate inevitably leads to the emergence of closed time-like curves as in the case of Taub-Nut. The existence of closed time-like curves violates strong and stable causality. Apparently the strongest causality condition of global hyperbolicity is not satisfied. Globally hyperbolic space-times admit a Cauchy surface the initial conditions on which determine all the events on the manifold. Technically, the domain of dependence of a Cauchy surface is the whole manifold. The boundary of the domain of dependence which is known as the Cauchy horizon is an empty set, for Cauchy surfaces. KTN and Taub-NUT space-times fail to satisfy global hyperbolicity as they do not admit a Cauchy surface.   To circumvent this problem, Bonnor suggested imposing Misner's identification partially and to treat the singularity at $\theta=\pi$ as a massless source of angular momentum \cite{bonnor}. The maximal extension of the KTN metric derived by Miller in \cite{miller} also applies to Bonnor's identification. Despite the fact that the KTN metric admits closed time-like curves, it provides a natural analytic tool  to probe the possible effects of gravitomagnetic monopoles. In that respect the perturbations and the thermodynamic properties of KTN space-time have been extensively studied \cite{ktn1,ktn2,ktn3,ktn4,ktn5,ktn6,ktn7,ktn8,ktn9,ktn10,ktn11}.

One can express the KTN metric in Kerr-like coordinates: 
\begin{eqnarray}
ds^2& =&\frac{1}{\Sigma}(\Delta -a^2\sin^2\theta)d t^2
    -\frac{2}{\Sigma}[\Delta A -a(\Sigma +a A)\sin^2\theta]d t
d \phi \nonumber \\ 
    &-&\frac{1}{\Sigma}[(\Sigma +a A)^2\sin^2\theta-A^2\Delta] d \phi^2
    -\frac{\Sigma}{\Delta}d r^2 -\Sigma d\theta^2
\label{metricktn}
\end{eqnarray}
where $m$ is the gravitational mass, $-\ell$ is the gravitomagnetic monopole moment or the NUT charge, and  $a=J/M$ is the Kerr parameter, where $J$ is the angular momentum parameter of the space-time. $\Sigma$, $\Delta$ and $A$ are defined by
\begin{eqnarray}
&\Sigma& = r^2 +(\ell +a \cos \theta)^2,\ 
\Delta = r^2-2Mr-\ell^2 + a^2 \nonumber \\ 
&A& = a \sin^2\theta -2\ell\cos\theta 
\end{eqnarray}
There exist an inner and an outer event horizon with spatial coordinates $r_{\pm}$, if the roots of $\Delta$ are real and positive. The metric (\ref{metricktn}) consists of three disjoint regions separated by Cauchy horizons. The regions $-\infty<r<r_-$ and $r_+<r<\infty$ constitute the Kerr-NUT region, whereas the region between the Cauchy horizons $r_- <r< r_+$ is called the Kerr-Taub region \cite{miller}. In the non-rotating case the corresponding Taub region --which is interpreted as a cosmological model-- is also separated from its extension (the NUT region) by a Cauchy horizon \cite{misnertaub}. The metric is non-singular in the three disjoint regions, except for the possible curvature singularity at $r=0$. The curvature scalars will not diverge if $a < \vert \ell \vert$, in which case $\Sigma$ cannot be zero. 

 The singularity theorems developed by Penrose and Hawking identify geodesic incompleteness with the existence of singularities \cite{singtheo}. Geodesic incompleteness refers to the in-extendibility of time-like and null geodesics beyond certain values of affine parameters. Physically this corresponds to freely falling observers disappearing off the edge of the Universe. One expects the curvature scalars to diverge to infinity at the singularities, though  no rigorous proof exists. Taub-NUT and Kerr-Taub-NUT space-times --which admit closed time-like curves-- constitute  counter-examples to this expectation. These space-times satisfy an idiosyncratic property known as imprisoned incompleteness. Incomplete time-like and null geodesics are totally imprisoned in compact neighbourhoods of the event horizon. The world-line of a freely falling observer continues to wind round and round in a compact set, never reaching beyond a certain value of the affine parameter (see e.g. \cite{hawkingellis}). Thus, the space-time is singular according to the definition of Penrose and Hawking based on geodesic incompleteness. However, the curvature scalars remain finite for $a < \vert \ell \vert$, contrary to the general expectation. 

On the other hand, $\Sigma$ can be zero for $a > \vert \ell \vert$ and there exists a curvature singularity at $r=0$ and $\theta=\arccos [(-\ell)/a]$, where $\Sigma=0$. In its weak form, the cosmic censorship conjecture proposed by Penrose asserts that the curvature singularities should be causally disconnected from the distant observers by the event horizons which serve as a one-way causal membrane \cite{ccc}.  An important feature of this conjecture is that if a solution admits curvature
singularities that are not covered by an event horizon, then small perturbations of that solution remove this property.

The condition for the existence of the event horizon for the KTN metric is:
\begin{equation}
\delta \equiv M^2 + \ell^2 -a^2 \geq 0
\label{condimain}
\end{equation}
where we have defined $\delta$. If $\delta$ is positive, there exist an inner and an outer event horizon at 
\begin{equation}
r_{\pm}=M \pm \sqrt{\delta}=M \pm \sqrt{M^2 + \ell^2 -a^2}
\end{equation}
The case $\delta=0$ corresponds to extremal black holes with $r_+=M$. However for $\delta < 0$, one cannot find a real root for $r_+$ which is the spatial location of the event horizon. In that case the space-time parameters describe a naked singularity which is not shrouded behind an event horizon. In Wald type problems, one starts with a black hole satisfying a main inequality in the form (\ref{condimain}), and perturbs it with test particles or fields. One checks whether the angular momentum or charge parameters could be increased beyond the extremal limit to make $\delta$ negative, which indicates that the final parameters of the space-time describe a naked singularity.  For example a Kerr black hole is said to be overspun by perturbations if $a^2>M^2$, and a Reissner-Nordstr\"{o}m black hole is said to be overcharged if $Q^2>M^2$. The spatial location of the event horizons are the roots of $\Delta$, which cannot be real when $\delta <0$. In that case the event horizon is said to be destroyed. For KTN black holes, overspinning occurs when $a^2>M^2 + \ell^2$ as implied by (\ref{condimain}). In this work we perturb a KTN black hole by test fields satisfying the null energy condition. Initially the parameters of the space-time satisfy (\ref{condimain}) with $\delta \geq 0$. After the interaction, the space-time settles to a new KTN solution with modified parameters. The energy and the angular momentum of the test field lead to  increases in the mass parameter and  angular momentum parameters respectively, while the NUT charge is left invariant in the interaction. Since the test fields do not carry a NUT charge themselves, they cannot change the NUT charge of the background space-time. The final value of $\delta$ defined in (\ref{condimain}) (i.e. $\delta_{\rm{fin}}$) is derived by substituting the final values of the parameters $M_{\rm{fin}}$, $a_{\rm{fin}}=(J_{\rm{fin}})/(M_{\rm{fin}})$ and $\ell_{\rm{fin}}=\ell$. If $\delta_{\rm{fin}}$ is negative at the end of the interaction, one cannot find a real root for the spatial coordinate of the event horizon. In that case the event horizon ceases to exist. The black hole is overspun into a naked singularity.  In his seminal work, Wald showed that perturbations carrying sufficiently large angular momentum or charge to destroy an extremal Kerr-Newman black hole, are not absorbed by the black hole \cite{wald74}. Following Wald many similar thought experiments were constructed involving particles \cite{hu,js,backhu,backjs,q1,q2,q3,q4,q5,q6,gao,q7,v1} and fields \cite{semiz,overspin,emccc,natario,duztas2,mode,tjphys}. In recent years the scope of research has been extended to higher dimensions, de-Sitter and anti de-Sitter space-times, and black holes in alternative theories of gravity \cite{magne,hong,sen3,kerrmog,btz,gwak3,chen,ongyao,ong,ghosh,mtz,ext1,jamil,shay3,shay4,he,he2,dilat,siahaan,kerrsen,yin,btz1,corelli,sia2,btzwill}
. It turns out that perturbations satisfying the null energy condition cannot destroy the event horizons of black holes, employing back-reaction effects if necessary. For these perturbations, there exists a lower limit for the energy to allow their absorption by the black hole.  When a test body or a field falls across the horizon the fluxes of energy, momentum and charge are related to the energy-momentum tensor of the test body/field by:
\begin{equation}
\delta M - \Omega \delta J - \Phi \delta Q= \int_{\rm{H}}(T_{\mu \nu} X^{\mu} X^{\nu}) dV
\end{equation}
where $X^{\mu}$ is null on the horizon. The null energy condition on the energy momentum tensor of the perturbation ($T_{\mu \nu} X^{\mu} X^{\nu} \geq 0$) implies that:
\begin{equation}
\delta M \geq \Omega \delta J + \Phi \delta Q
\label{needham}
\end{equation}
where $\delta M$ is the energy, $\delta J$ is the angular momentum, and $\delta Q$ is the charge of the perturbation, whereas $\Omega$ is the angular velocity of the event horizon and $\Phi$ is the electrostatic potential for charged black holes. In Kerr family of metrics the angular velocity of the event horizon is given by $\Omega=-(g_{t\phi})/(g_{\phi \phi})$ evaluated at the horizon $r=r_+$, which corresponds to the angular velocity that the observers in the ergosphere are forced to rotate. The condition (\ref{needham}) first derived by Needham \cite{needham}, determines the lower limit for the energy of a perturbation ($\delta M$) to allow its absorption by the black hole.   Needham's condition (\ref{needham}) applies to test bodies and fields the energy-momentum tensor of which satisfy the null energy condition. The energy and angular momentum parameters of  test fields are related by
\begin{equation}
\delta J = \frac{m}{\omega} \delta M
\label{jmratio}
\end{equation}
where $m$ is the azimuthal number and $\omega$ is the frequency. As $\omega$ decreases, the relative contribution of test fields to the angular momentum parameter increases. Therefore the low frequency modes lead to a relatively higher increase in the angular momentum parameter, challenging the validity of the main inequality (\ref{condimain}).  In that respect, we will refer to the low frequency modes as the challenging modes. From (\ref{jmratio}), one could infer that the modes with arbitrarily low frequencies would destroy the event horizon by increasing the angular momentum beyond the extremal limit. However  the absorption of the modes with frequencies $\omega < m\Omega$ is not allowed, as implied by the Needham's condition (\ref{needham}). These modes are reflected back to infinity with a larger amplitude, in the well-known effect of superradiance. The superradiance condition is derived by analysing the corresponding wave equation for test fields  in the curved background of black hole space-times. It occurs for integer spin test fields satisfying the null energy condition. Superradiance occurs for test field modes with frequencies  below the critical value $\omega=m \Omega$, for which  the reflection coefficient is larger than unity and the transmission coefficient is negative. Therefore no net absorption of these modes occurs as they are reflected back to infinity with a larger amplitude. This is the natural mechanism which precludes the violation of the main equality (\ref{condimain}) or its analogues, thus preventing the destruction of the event horizon. However, a corresponding lower limit does not exist for fermionic fields which do not satisfy the null energy condition. When the null energy condition is not satisfied , one cannot derive  Needham's condition (\ref{needham}). However, one cannot directly infer that a lower limit does not exist for fermions. This only renders us inconclusive whether there exists a lower limit for fermions. In \cite{bini}, Bini et \textit{al.} studied the wave equations for test fields in KTN space-time. They derived that superradiance occurs for bosonic fields and it is absent for fermionic fields $s=1/2$ and $s=3/2$ analogous to the Kerr case. The absorption of the low energy modes is allowed which leads to a generic destruction of the event horizon \cite{duztas,toth,generic,spinhalf,threehalves}. Second order variations which naturally contribute to the interaction to second order, cannot fix this problem. The solution of this problem appears to be postponed to a quantum theory of gravity.

Previously we have evaluated the possibility to overspin a KTN black hole by sending in test fields from infinity \cite{taubnut}. The analysis in that work is restricted to the contributions of the first order perturbations $(\delta M)$ and $(\delta J)$. The analysis has been carried out to second order involving $(\delta M)^2$ and $(\delta J)^2$ terms. However the contribution of the second order variations $\delta^2 M$ and $\delta^2 J$ has not been incorporated. We derived that overspinning is possible, however it is not generic in the sense that it occurs by fine-tuned parameters of the perturbations and the final value of $\delta$ defined in (\ref{condimain})  is second order in the perturbations; i.e. $\delta_{\rm{fin}} \sim -(\delta M)^2$. This strongly suggests the destruction of the event horizon ($\delta <0$) is just an artefact due to neglecting the second order contributions by the second order variations $\delta^2 M$ and $\delta^2 J$. Over-spinning is likely to be fixed by incorporating the contribution of the second order variations. 

The incorporation of second order variations has been notoriously challenging in Wald type problems. Recently, Sorce and Wald derived the second order analogue of Needham's condition which could serve as a systematic method to incorporate the second order contributions \cite{w2}.
\begin{equation}
\delta^2 M-\Omega \delta^2 J-\Phi \delta^2 Q \geq  \frac{\kappa}{8\pi}\delta^2 A
\label{condisorcewald}
\end{equation}
where $\kappa$ is the surface gravity and $A$ is the area of the event horizon
. In \cite{spin2} we have demonstrated how Sorce-Wald condition (\ref{condisorcewald}) can be correctly used to fix the overspinning and overcharging problems of Kerr and Reissner-Nordstr\"{o}m black holes, previously studied by Hubeny \cite{hu}, Jacobson-Sotiriou \cite{js}, and D\"{u}zta\c{s}-Semiz \cite{overspin}.

In this work we re-evaluate the overspinning problem for Kerr-Taub-NUT black holes. 
 In Sec. (\ref{first}) we use  Needham's condition (\ref{needham})  and apply a second order analysis which involves $(\delta M)^2$ and $(\delta J)^2$ terms, but neglects the second order variations $\delta^2 M$ and $\delta^2 J$. We re-derive the result in \cite{taubnut}: $\delta_{\rm{fin}} =-M^2 \epsilon^2= -(\delta M)^2$. In Sec. (\ref{second}), we incorporate the contribution of the second order variations using the Sorce-Wald condition (\ref{condisorcewald}). We demonstrate how the second order variations modify $\delta_{\rm{fin}}$ to make it positive. 

\section{Test fields and first order variations}
\label{first}
Previously, we derived that test fields satisfying the null energy condition can overspin Kerr-Taub-NUT black holes into naked singularities \cite{taubnut}. The analysis in this work is carried out to second order, involving the second order terms $(\delta M)^2$ and $(\delta J)^2$. However the contribution of the second order perturbations is ignored which renders the analysis incomplete. In this section we re-evaluate this problem with some amendments to reveal how over-spinning occurs when one ignores the second order variations. We incorporate the second order variations in the next section,  anticipating to find the same result as in the Kerr-Newman case; i.e. the perturbations satisfying the null energy condition cannot destroy event horizons.  

We start with a KTN black hole with a curvature singularity at $r=0$ and $\theta=\arccos [(-\ell)/a]$, which implies $a>\vert \ell \vert$. We parametrize the KTN black hole:
\begin{equation}
\delta_{\rm{in}}= M^2 +\ell^2 -\frac{J^2}{M^2}=M^2\epsilon^2
\label{param1}
\end{equation}
where $\epsilon \ll 0$ parametrizes the closeness to extremality. The initial value of the $\delta$ function  implies $r_+ =M(1+\epsilon)$, initially. We define the dimensionless parameters as we did in \cite{taubnut}
\begin{equation}
\beta \equiv \frac{\vert \ell \vert}{M}, \quad \alpha \equiv \frac{J}{M^2}=\frac{a}{M}
\label{dimless}
\end{equation}
where $\beta$ is the dimensionless NUT charge and $\alpha$ is the dimensionless angular momentum parameter. We re-write the parametrization (\ref{param1}) in terms of the dimensionless variables.
\begin{equation}
1+\beta^2 -\alpha^2 =\epsilon^2
\label{paramdimless}
\end{equation}
We envisage a test field with frequency $\omega$ and azimuthal number $m$, incident on the black hole from infinity. We choose the energy and the angular momentum parameters of the test field as:
\begin{equation}
\delta M=M\epsilon, \quad \delta J=\frac{m}{\omega}\delta M=\frac{m}{\omega}M\epsilon
\label{deltamdeltaj}
\end{equation}
in accord with the test field approximation. We re-write Needham's condition (\ref{needham}) for KTN black holes.
\begin{equation}
\delta M \geq \Omega \delta J  
\label{needhamktn}
\end{equation}
Note that, for test fields satisfying (\ref{deltamdeltaj}), the condition (\ref{needhamktn}) reduces to the well-known superradiance condition. By direct substitution one finds:
\begin{equation}
\omega \geq m\Omega
\end{equation}
In \cite{bini} Bini \textit{et al.} separated the wave equations for spin-0, spin-1/2, spin-1, spin-3/2, and spin-2 fields in KTN space-time, following the seminal work by Teukolsky \cite{teuk} for Kerr black holes. They evaluated the transmission-reflection coefficients for the radial part of  the wave equations for  scalar fields and derived that the reflection coefficient is larger than 1; i.e. superradiance occurs for the modes whose frequency is less than:
\begin{equation}
\omega<\omega_{\rm{sl}}=\frac{ma}{2(Mr_+ + \ell^2)}=m\Omega_{\rm{KTN}}
\label{omegasl}
\end{equation}
 where $\omega_{\rm{sl}}$ denotes the super-radiance limit and $\Omega_{\rm{KTN}}$ is angular velocity of the event horizon for KTN black holes. (We proceed by dropping  the subscript (KTN))  
The test fields with frequencies below the super-radiance limit are not absorbed by the black hole, but reflected back to infinity with a larger amplitude. No net absorption of these test fields occurs. To analyse the over-spinning problem we first demand that the test field is absorbed by the black hole; i.e. the frequency satisfies
\begin{equation}
\omega \geq \omega_{\rm{sl}}=\frac{ma}{2(Mr_+ + \ell^2)}=\frac{m\alpha}{2M(\alpha^2 +\epsilon^2 +\epsilon)}
\label{omegamin}
\end{equation}
Note that $r_+=M(1+\epsilon)$ for a KTN black hole parametrized as (\ref{param1}). We have expressed the  angular velocity of the event horizon and the super-radiance limit in terms of the dimensionless variables defined in (\ref{dimless}). 

Next we demand that the test field overspins the KTN black hole into a naked singularity. Equivalently we demand that the final parameters of the space-time satisfy:
\begin{equation}
\delta_{\rm{fin}} = (M+ \delta M)^2 + \ell^2 -\frac{(J+ \delta J)^2}{(M+\delta M)^2} <0
\label{deltafinneg}
\end{equation}
We can express the condition (\ref{deltafinneg}) in terms of the dimensionless variables defined in (\ref{paramdimless}) and proceed to derive that (\ref{deltafinneg}) is satisfied and the event horizon can be destroyed if the frequency of the test field satisfies (see \cite{taubnut})
\begin{equation}
\omega < \frac{m\epsilon}{M[(1+\epsilon)\sqrt{\alpha^2 +2\epsilon^2 +2\epsilon}-\alpha]} \equiv \omega_{\rm{max}}
\label{omegamax}
\end{equation}
The maximum value of the frequency is derived in \cite{taubnut} by imposing that a test field described in (\ref{deltamdeltaj}), makes the expression in (\ref{deltafinneg}) negative at the end of the interaction. For a fixed value of $\delta M$, the frequency $\omega$ and $\delta J$ are inversely proportional, which is manifest in (\ref{deltamdeltaj}). As $\omega$ decreases, $\delta J$ increases, and below a certain value  $\omega=\omega_{\rm{max}}$, the expression in (\ref{deltafinneg}) becomes negative. However one cannot directly conclude that the black hole has been overspun. The modes with frequencies less than $\omega_{\rm{sl}}$ are not absorbed by the black hole due to superradiance. Therefore we first demand that the frequency of the test field is larger than $\omega_{\rm{sl}}$. As we mentioned the low frequency modes are more challenging as their contribution ($\delta J$) to the angular momentum parameter  is larger. Below a certain value --which is is derived in (\ref{omegamax})-- ($\delta J$) will be sufficiently large to make the right-hand side of (\ref{deltafinneg}) negative. However, if this value is below the superradiance  limit, the test field will not be absorbed by the black hole. Therefore the frequency should also be larger than the superradiance limit (\ref{omegamin}). For overspinning to occur the conditions (\ref{omegamin}) and (\ref{omegamax}) should be satisfied simultaneously. A low energy mode with a frequency lower than $\omega_{\rm{max}}$ defined in (\ref{omegamax}), should also be absorbed by the black hole, which entails that  $\omega_{\rm{max}}$ should be larger than the limiting frequency for superradiance. In (\cite{taubnut}), we expressed the super-radiance limit $\omega_{\rm{sl}}$ and the upper limit $\omega_{\rm{max}}$ in terms of the dimensionless parameters as given in (\ref{omegamin}) and (\ref{omegamax}). We demanded that $\omega_{\rm{max}}>\omega_{\rm{sl}}$ so that the modes that make the right hand side of (\ref{deltafinneg}) negative are absorbed by the black hole. We evaluated the inequality $\omega_{\rm{max}}>\omega_{\rm{sl}}$ in terms of the dimensionless parameters, to second order in $\epsilon$.    We derived that the condition $\omega_{\rm{max}}>\omega_{\rm{sl}}$ is equivalent to:
\begin{equation}
\alpha^2 >\frac{2+2\epsilon}{2+3\epsilon}
\label{alphacrit}
\end{equation}
The condition (\ref{alphacrit}) implies that $\alpha \sim 1$ for overspinning to occur i.e. the initial angular momentum parameter is almost as large as the mass parameter. For a generic KTN black hole which does not satisfy (\ref{alphacrit}), $\omega_{\rm{max}}<\omega_{\rm{sl}}$. Therefore the modes that can make $\delta_{\rm{fin}}$ in (\ref{deltafinneg}) negative, are reflected back to infinity due to super-radiance. Overspinning can only occur for the sub-collection of KTN black holes that satisfy (\ref{alphacrit}), and by test fields with frequency in the range $\omega_{\rm{sl}}<\omega<\omega_{\rm{max}}$. This range only exists for the sub-collection of KTN black holes that satisfy (\ref{alphacrit}), or equivalently $\omega_{\rm{max}}>\omega_{\rm{sl}}$. Test field modes with frequencies in this range  both get absorbed by the black hole ($\omega>\omega_{\rm{sl}}$) and make $\delta_{\rm{fin}}$ negative ($\omega<\omega_{\rm{max}}$), ignoring the contribution of the second order perturbations. 

We should also demand that the NUT charge is non-zero; i.e. $\beta^2 >0$. This part was ignored in the original work \cite{taubnut}. The initial state of the KTN black hole in terms of the dimensionless parameters given in (\ref{paramdimless}) implies
\begin{equation}
\alpha^2 +\epsilon^2 -1 >0
\label{alphacrit2}
\end{equation}
Therefore overspinning can occur for KTN black holes with dimensionless parameters:
\begin{equation}
\alpha^2=1-\eta^2, \quad \beta^2=\epsilon^2 - \eta^2
\label{etaalphabeta}
\end{equation}
where $\eta^2 < \epsilon^2 \ll 0$ to ensure that $\beta >0$. Note that $\alpha$ and $\beta$ are not necessarily small parameters. However for KTN black holes that can possibly be overspun by test fields (that satisfies (\ref{alphacrit})),  the dimensionless NUT charge $\beta$ should be very small and the dimensionless angular momentum parameter $\alpha$ should be arbitrarily close to $1$. These stringent conditions on the initial values of the  space-time parameters and the narrow range of frequencies $m\Omega <\omega <\omega_{\rm{\max}}$ for the test field, indicate that the overspinning of KTN black holes is not generic. To evaluate whether the over-spinning is generic, let us consider the most challenging  mode with $\omega=m\Omega$. The relative contribution of the modes to the angular momentum parameter increases as the frequency $\omega$ decreases. One could infer that the modes with lower frequencies would be more  likely to lead to a generic over-spinning. However, no net absorption of the modes with frequencies lower than $\omega_{\rm{sl}}=m\Omega$ due to super-radiance or equivalently Needham's condition. Therefore we exclude these modes in our analysis for over-spinning. We start with a KTN black hole parametrized as (\ref{param1}) with $\alpha=1$ and $\beta^2=\epsilon^2$; i.e. the dimensionless NUT charge is of the order of $\epsilon$ which parametrizes the closeness to extremality. In \cite{taubnut} we showed that there is no possibility for over-spinning to occur unless the NUT charge is drastically small. We send in a test field from infinity with frequency $\omega=m\Omega$, and energy $\delta M=M\epsilon$. The contribution of this test field to the angular momentum parameter is:
\begin{equation}
\delta J=\frac{m}{\omega} \delta M=\frac{M\epsilon}{\Omega}=2M^2(\epsilon + \epsilon^2)
\label{deltajtest}
\end{equation} 
The final value of the indicator $\delta$ is given by:
\begin{eqnarray}
\delta_{\rm{fin}} &=& (M_{\rm{fin}})^2 -\frac{(J_{\rm{fin}})^2}{(M_{\rm{fin}})^2} + \ell_{\rm{fin}}^2  \nonumber \\
&=& M^2(1+\epsilon)^2 - \frac{\left( J+2M^2(\epsilon+\epsilon^2) \right)^2}{M^2(1+\epsilon)^2} +\ell^2 \nonumber \\
&=& M^2(1+\epsilon)^2 -\frac{J^2}{M^2(1+\epsilon)^2}-\frac{4J(\epsilon+\epsilon^2)}{(1+\epsilon)^2} \nonumber \\
&-&\frac{4M^2\epsilon^2}{(1+\epsilon)^2} +\ell^2
\end{eqnarray}   
where the initial values of the space-time parameters $M,J,\ell$ satisfy (\ref{param1}). Note that to second order
\[
\frac{1}{(1+\epsilon)^2}=(1-2\epsilon +3\epsilon^2) 
\]
which leads to
\begin{eqnarray}
\delta_{\rm{fin}} &=& 2M^2 \epsilon^2 +2M^2\epsilon +2\frac{J^2}{M^2}\epsilon -3\frac{J^2}{M^2}\epsilon^2 \nonumber \\
&-& 4J\epsilon +4J \epsilon^2 -4M^2\epsilon^2 
\end{eqnarray} 
We also substitute
\begin{equation}
J^2=M^4\alpha^2=M^4(1-\eta^2) ; J=M^2(1-\eta^2/2)
\label{deltajmeta}
\end{equation}
However, $\eta^2$ terms contribute to third and fourth order when multiplied by $\epsilon$ and $\epsilon^2$. To second order the final value of $\delta$ is calculated as:
\begin{equation}
\delta_{\rm{fin}} =-M^2 \epsilon^2
\label{deltafinfirst}
\end{equation}
Note that we have started with a KTN black hole with $r_+=M+\sqrt{\delta_{\rm{in}}}=M(1+\epsilon)$. At the end of the interaction, the negative sign for $\delta_{\rm{fin}}$ indicates that the event horizon ceases to exist since $r_+=M+\sqrt{\delta_{\rm{fin}}}$ has no real roots. One may conclude that the final parameters of the space-time describe a naked singularity. However, the magnitude of $\delta_{\rm{fin}}$ is second order which suggests the incorporation of the second variations can fix the over-spinning problem to make $\delta_{\rm{fin}}$ positive again.
\section{Sorce-Wald condition and second order variations}
\label{second}
The Sorce-Wald condition (\ref{condisorcewald}) derived in \cite{w2} provides a systematic tool to evaluate the effect of second order variations. In this section we incorporate the Sorce-Wald condition into the calculation of $\delta_{\rm{fin}}$ to achieve a complete second order analysis. For an electrically neutral, rotating black hole the Sorce-Wald condition reduces to:
\begin{equation}
\delta^2 M-\Omega \delta^2 J \geq  \frac{\kappa}{8\pi}\delta^2 A
\label{condisorcewaldktn}
\end{equation}
The condition is evaluated for a KTN black hole in the Appendix, which leads to
\begin{equation}
\delta^2 M-\Omega \delta^2 J \geq \frac{(\delta J)^2}{4M^3 (1+\epsilon+\beta^2)}=\frac{(\delta J)^2}{4M^3} +O(3)
\label{sorcewaldresult1}
\end{equation} 
where the KTN black hole is parametrised as (\ref{param1}). Note that $(\delta J)^2$ is inherently a second order quantity for a test particle or field, and $\beta^2$ is second order for a KTN black hole that can possibly be overspun (that satisfies (\ref{alphacrit})). We re-write $\delta_{\rm{fin}}$ with second order perturbations:
\begin{eqnarray}
\delta_{\rm{fin}}&=& M_{\rm{fin}}^2+ \ell^2 -\frac{J_{\rm{fin}}^2}{M_{\rm{fin}}^2} \nonumber \\
&=&(M+ \delta M +\delta^2 M)^2 + \ell^2 -\frac{(J+ \delta J+ \delta^2 J)^2}{(M+\delta M+\delta^2 M)^2} \nonumber \\
&&
\end{eqnarray}
We expand $\delta_{\rm{fin}}$ to second order:
\begin{eqnarray}
\delta_{\rm{fin}}&=& M^2 +(\delta M)^2 +2M (\delta M) +2M (\delta ^2 M) +\ell^2 \nonumber \\
&-& \frac{J^2 +(\delta J)^2 +2J (\delta J) +2J (\delta ^2 J)}{M^2 +(\delta M)^2 +2M (\delta M) +2M (\delta ^2 M)}
\end{eqnarray}
We substitute $(\delta M)=M\epsilon$ for a test field. Note that 
\begin{widetext}
\begin{equation}
\frac{J^2 +(\delta J)^2 +2J (\delta J) +2J (\delta ^2 J)}{M^2 \left( 1+ \epsilon^2 + 2\epsilon +\frac{2(\delta ^2 M)}{M} \right) }=\frac{J^2 +(\delta J)^2 +2J (\delta J) +2J (\delta ^2 J)}{M^2}\left(1-2\epsilon+3\epsilon^2 - \frac{2(\delta ^2 M)}{M} \right)
\end{equation}
\end{widetext}
Using the result in the previous section (\ref{deltafinfirst}), one finds that:
\begin{equation}
\delta_{\rm{fin}}=-M^2\epsilon^2 +2M \delta^2 M  +\frac{2J^2}{M^3} \delta^2 M - \frac{2J}{M^2} \delta^2 J
\end{equation}
Using (\ref{deltajmeta}):
\begin{equation}
\delta_{\rm{fin}}=-M^2\epsilon^2 + 4M \delta^2 M -2 \delta^2 J
\label{deltafinsecond}
\end{equation}
The second order variations have modified the final value of $\delta$. We are going to use the Sorce-Wald condition to check if the modification due to the second order variations can render the final value of $\delta$ positive. Note that 
\begin{equation}
\Omega \delta^2 J= \frac{\alpha}{2M(\alpha+ \epsilon +\epsilon^2)}=\frac{\delta^2 J}{2M} +O(3)
\end{equation}
Also using (\ref{deltajtest})
\begin{equation}
(\delta J)^2=4M^4 (\epsilon +\epsilon^2)^2=4M^4\epsilon^2 + O(3)
\end{equation}
We re-write the Sorce-Wald condition for KTN black holes in the form:
\begin{equation}
2M \delta^2 M - \delta^2 J \geq 2M^2 \epsilon^2
\label{sorcewaldfinal}
\end{equation}
By direct substitution of the Sorce-Wald condition (\ref{sorcewaldfinal}) into (\ref{deltafinsecond}), we derive that
\begin{equation}
\delta_{\rm{fin}}\geq 3M^2\epsilon^2
\end{equation}
When we execute an analysis incorporating the second order variations, the final value of $\delta$ becomes positive. There exists an event horizon at 
\begin{equation}
r_+=M+ \sqrt{\delta_{\rm{fin}}} \sim M(1+\sqrt{3}\epsilon)
\end{equation}
The final parameters of the space-time represent a black hole surrounded by an event horizon at $r_+\sim M(1+\sqrt{3}\epsilon)$, rather than a naked singularity. 
\section{Conclusions}
Previously, we had shown that Kerr-Taub-NUT black holes can be overspun into naked singularities by test fields \cite{taubnut}. The frequency of the test field was restricted to a narrow range bounded below by the super-radiance limit, as expected in attempts to overspin black holes with test fields satisfying the null energy condition. The analysis was carried out to second order involving the effects of $(\delta M)^2$ and $(\delta J)^2$ terms. The effect of the second order variations $\delta^2 M$ and $\delta^2 J$ was ignored.  The final state of the space-time describes a naked singularity or a black hole surrounded by an event horizon depending on the sign of the $\delta$ function defined in (\ref{condimain}). In \cite{taubnut} we derived that the final value of $\delta$ is second order in magnitude; i.e. $\delta_{\rm{fin}} \sim -M^2 \epsilon^2 $. This suggests that a complete second order analysis involving the effect of second order variations could lead to a positive value for $\delta_{\rm{fin}}$.

The effect of second order variations can be systematically incorporated into the expression for $\delta_{\rm{fin}}$ by the Sorce-Wald condition (\ref{condisorcewald}) derived in \cite{w2}. However Sorce-Wald and the subsequent authors adapt an alternative method in the final stage. Instead of  $\delta_{\rm{fin}}$ used by Semiz \cite{semiz} and this author, they define a function of an additional small parameter $(f(\lambda))$ and expand the function to second order. In \cite{absorp} and \cite{spin2} we showed that the introduction of the additional small parameter leads to order of magnitude problems. The argument is simple. In the test particle/field approximation $(\delta M)$ and $(\delta J)$ are inherently first order, whereas $(\delta M)^2$ and $(\delta J)^2$ and the second order variations $(\delta^2 M)$ and $(\delta^2 J)$ are second order quantities. It is an algebraic fact that $\lambda^2 (\delta^2 M)$ is a fourth order quantity, therefore the lowest order contribution of the second order variations to $f(\lambda)$ is fourth order (not second). We pedagogically elucidated this fact in \cite{spin2}, and also demonstrated how the Sorce-Wald condition can be properly incorporated into the analysis by the line of research developed by Semiz \cite{semiz} and this author. We evaluated the Sorce-Wald condition for Kerr and Reissner-Nordst\"{o}m black holes, and calculated 
$\delta_{\rm{fin}}$ with second order variations. The second order variations contribute to $\delta_{\rm{fin}}$ to second order --as they should-- and modify the value of $\delta_{\rm{fin}}$ to make it positive.

Here we re-evaluated the interaction of nearly extremal  KTN black holes with test fields. In Sec. (\ref{first}), we re-derived and extended our previous results in \cite{taubnut}. Even if one ignores the second order variations, over-spinning is not a generic possibility for KTN black holes. The event horizon can be destroyed for a sub-collection of KTN black holes whose spin parameter $a$ is arbitrarily close to the mass parameter $M$, which entails that the NUT charge $\ell$ is arbitrarily small as the black hole is close to extremality. The conditions satisfied by KTN black holes that can be overspun by test fields when the second order variations are ignored, are made precise in  (\ref{alphacrit}), (\ref{alphacrit2}) ,and (\ref{etaalphabeta}) in terms of the dimensionless parameters defined in (\ref{dimless}). The super-radiance limit $w=m\Omega$, determines the most challenging modes to destroy the event horizon. The lower frequency modes which would give a larger contribution to the spin parameter, are not absorbed by the black hole. Therefore they are excluded from the analysis. When one perturbs a nearly extremal KTN black hole satisfying the above-mentioned properties (that can possibly be overspun) with optimal test fields ($w=m\Omega$) and ignores the second order variations, the final value of the $\delta$ function can be negative albeit second order in magnitude i.e. $\delta_{\rm{fin}}=-M^2 \epsilon^2$. In Sec. (\ref{second}), we incorporated the effect of the second order variations, using the Sorce-Wald condition. We evaluated the Sorce-Wald condition for a nearly extremal KTN black hole whose spin parameter $a$ is arbitrarily close to its mass parameter $M$, and the NUT charge $\ell$ is arbitrarily small. We also used the most challenging modes which have the lowest possible frequency $\omega=m\Omega$ that can be absorbed by the black hole. The Sorce-Wald condition involves the second variation of the area, which is a rather tedious calculation given in the Appendix.  We carried out a calculation to second order and evaluated the modified value of $\delta_{\rm{fin}}$ using the Sorce-Wald condition. We have avoided the common order of magnitude problems encountered in \cite{w2} (Sorce-Wald) and subsequent works where the second order variations appear as a product with $\lambda^2$. In our calculation, the contribution of the second order variations to $\delta_{\rm{fin}}$ is actually second order We derived that $\delta_{\rm{fin}}\geq 3M^2\epsilon^2$, when one incorporates the second order variations. There exists an event horizon at $r_+\sim M(1+\sqrt{3}\epsilon)$. Not only the event horizon has been preserved, but also its spatial location $r_+$ has increased, which implies the area of the black hole has increased $A_{\rm{fin}}>A_{\rm{in}}$. The second law of black hole dynamics identifies the area of the event horizon with entropy which entails that the area should not decrease \cite{hawkingarea}. The third law asserts that the surface gravity cannot be reduced to zero in any continuous process, which implies that a black hole cannot be driven to extremality $(\delta_{\rm{fin}}=0)$ \cite{israelthirdlaw}. The fact that we started with a KTN black hole satisfying $\delta_{\rm{in}}= M^2\epsilon^2$ and ended up with $\delta_{\rm{fin}}\sim 3M^2 \epsilon^2$ implies that the black hole has been driven away from extremality.  The two results --$A_{\rm{fin}}>A_{\rm{in}}$ and $\delta_{\rm{fin}}>\delta_{\rm{in}}$-- are in accord with the second and the third laws of black hole dynamics respectively, which apply to perturbations satisfying the null energy condition.

\section*{Appendix: Sorce-Wald condition for KTN black holes}
In this section we evaluate the Sorce-Wald condition (\ref{condisorcewald}) for a KTN black hole parametrized  as (\ref{param1}). Note that the surface area of a KTN black hole is
\begin{equation}
A=4\pi (r_+^2+a^2+\ell^2)
\end{equation}
The parametrization (\ref{param1}) implies that $r_+=M(1+\epsilon)$, which leads to:
\begin{equation}
A=4\pi \left(M^2(1+\epsilon)^2+\frac{J^2}{M^2}+\ell^2 \right)
\end{equation}
Note that there is no variation in $\ell$ due to the perturbations. Therefore $\delta^2 A$ for KTN black holes has the same structure as Kerr (see Equation (113) in \cite{w2}. The Kerr case; namely the $Q \to 0$ limit is evaluated in \cite{spin2}). The second order variation in the area of the black hole is given by :
\begin{widetext}
\begin{eqnarray}
\delta^2 A&=& -\frac{8\pi}{M^8 \epsilon^3} \left \{ (\delta M)^2 \left[ J^4 + (2+\epsilon^2)J^2 M^4 -M^8(1+\epsilon)(-1 +\epsilon +2\epsilon^2) \right]  \right. \nonumber \\
&+&   (\delta J)^2 \left[J^2M^2 + M^6 \epsilon^2 \right ] 
+ (\delta M \delta J ) \left[ -2J^3M-2JM^5(1+\epsilon^2) \right] \left. \right\}
\label{areavariation}
\end{eqnarray}
\end{widetext}

We evaluate the Sorce-Wald condition for KTN black holes  and for optimal perturbations which saturate Needham's condition:
\begin{equation}
\delta M=\Omega \delta J
\end{equation}
Recall (\ref{omegasl}), and (\ref{omegamin}) for the angular velocity of the event horizon.
\begin{widetext}
\[
\Omega=\frac{(\alpha)}{2M(1+\epsilon +\beta^2)}=\frac{1}{2M}\left( 1-\frac{\eta^2}{2} \right) \left( 1-\epsilon -\epsilon^2 -\beta^2 \right)=\frac{1}{2M} \left( 1-\epsilon -\epsilon^2 -\beta^2 -\frac{\eta^2}{2} \right) +O(3)
\]
\end{widetext}
which leads to
\[
\Omega^2 =\frac{1}{4M^2}(1-2\epsilon -\epsilon^2 -2\beta^2 -\eta^2 )
\]
where we have used (\ref{etaalphabeta}) for the dimensionless parameters. (Note that for small $\eta$, $\alpha^2=1-\eta^2$ implies $\alpha=1-\eta^2 /2 +O(3)$. ) Also note that $J^2=M^4(1-\eta^2)$ implies
\[
J=M^2 \left(1-\frac{\eta^2}{2} \right) \quad J^3= M^6 \left(1-\frac{3}{2} \eta^2 \right)
\]
With these substitutions, we first evaluate $(\delta M)^2$ terms in (\ref{areavariation}). After some algebra , we get
\[
(\delta M)^2 4M^8 \left( 1-\eta^2 -\frac{\epsilon^2}{2} \right)
\]
Substituting $(\delta M)^2=\Omega^2 (\delta J)^2$
\begin{equation}
(\delta J)^2 M^6 \left(1- 2\epsilon -\frac{3}{2} \epsilon^2 -2\beta^2 -2\eta^2 \right)
\label{deltamsquare}
\end{equation}
Next, we evaluate $(\delta J)^2$ terms:
\begin{equation}
(\delta J)^2 (J^2M^2 + M^6\epsilon^2)=(\delta J)^2  M^6 (1+\epsilon^2 -\eta^2)
\label{deltajsquare}
\end{equation}
We proceed with $(\delta M) (\delta J)$ terms. First note that:
\[
-2(\delta M)(\delta J) [J^3 M + JM^5(1+\epsilon^2)]=2M^7 (\delta M)(\delta J) (-2 -\epsilon^2 +\eta^2)
\]
Substituting $(\delta M)=\Omega (\delta J)$
\begin{equation}
(\delta J)M^6 (-2+2\epsilon +\epsilon^2 +2\beta^2 +3\eta^2)
\label{deltamdeltajyarea}
\end{equation}
Adding the three contributions (\ref{deltamsquare}), (\ref{deltajsquare}) and (\ref{deltamdeltajyarea}), we derive the second variation in the area of a KTN black hole. 
\begin{equation}
\delta^2 A =-\frac{8\pi}{M^8 \epsilon^3}\left \{ (\delta J)^2 M^6 \frac{\epsilon^2}{2} \right\}
\end{equation}
To evaluate the Sorce-Wald condition for KTN black holes, we express the surface gravity in terms of the dimensionless parameters:
\[
\kappa=\frac{r_+ - r_-}{2(r_+^2 +a^2 +\ell^2)}=\frac{\epsilon}{2M(1+\epsilon +\beta^2)}
\]
Substituting the expressions for $\delta^2 A$ and $\kappa$ in (\ref{condisorcewaldktn}), we derive
\begin{equation}
\delta^2 M - \Omega \delta^2 J \geq \frac{(\delta J)^2}{4M^3(1+\epsilon+\beta^2)}
\label{sorcewaldcondiktn}
\end{equation}

\end{document}